\documentclass[journal]{IEEEtran}
\def\ps@headings{
\def\@oddhead{\mbox{}\scriptsize\rightmark \hfil \thepage}
\def\@evenhead{\scriptsize\thepage \hfil \leftmark\mbox{}}
\def\@oddfoot{}
\def\@evenfoot{}}
\makeatother
\usepackage{cite}
\usepackage{subfigure}
\usepackage{graphicx}
\usepackage{epstopdf}
\usepackage{url}
\usepackage{balance}
\begin{document}
\title{A General, Tractable and Accurate Model for a Cascade of Caches}
\author{G. Bianchi,  N. Blefari Melazzi, A. Caponi, A. Detti\\
Universit\`a degli Studi di Roma - Tor Vergata / CNIT, Italy \\
name.surname@uniroma2.it
}
\markboth{\tiny This work has been submitted to the IEEE for possible
publication. Copyright may be transferred without notice, after
which this version may no longer be accessible.}{}
\pubid{~}
\maketitle

\begin{abstract}
Performance evaluation of caching systems is an old and widely investigated research topic. The research community is once again actively working on this topic because the Internet is evolving towards new transfer modes, which envisage to cache both contents and instructions within the network. In particular, there is interest in characterizing multi-cache systems, in which requests not satisfied by a cache are forwarded to other caches.

In this field, this paper contributes as follows. First, we devise a simple but accurate approximate analysis for caches fed by general ``renewal'' traffic patterns. Second, we characterize and model the traffic statistics for the output (miss) stream. Third, we show in the simple example case of tandem caches how the resulting output stream model can be conveniently exploited to analyze the performance of subsequent cache stages. The main novelty of our work stems in the ability to handle traffic patterns beyond the traditional independent reference model, thus permitting simple assessment of cascade of caches as well as improved understanding of the phenomena involved in cache hierarchies.

\end{abstract}

\section{Introduction}

Caching is a technique used to temporarily store data, usually coming from an origin source, within a memory quickly accessible from the intended user of that data. Caches decrease access time and/or system load, as repeated requests for the same data are served by a fast/local memory rather than by a slower/remote source.

Requests of a data item to a single cache system result in a cache hit when that item is found in the cache. In case of a cache miss, the request is forwarded to the origin location of the item, if any. In multi-cache systems, requests not served by a cache are forwarded to other caches, before eventually reaching the origin source. Multi-cache systems can have different topologies, including cascade or hierarchical configurations.

When caches are full, a replacement policy chooses which item is to be cancelled to make room for a new item, following a cache miss. A popular replacement policy is the Least Recently Used (LRU) algorithm, which discards the least recently used data item first.

We are interested in studying the performance of networks of caches operating with the LRU policy.
The motivation for a renewed interest in cache networks lies in the proposed evolution of the Internet towards a so called Information Centric Network (ICN) \cite{Kop07,Jac09,Ahl12,Xyl13}. An ICN provides users with contents exposed as names, instead of providing communication channels between hosts; the network transfers individual, identifiable content chunks, instead of unidentifiable data containers (i.e., IP packets). Content chunks are explicitly addressed by-name allowing to perform content caching systematically and on-the-fly, potentially in every network node, and without the need to deploy cumbersome tasks such as HTTP header parsing \cite{Gho11,Sal12}. In ICN, all Internet users, irrespective of size or status may have their content cached. In addition, proposed architectures such as \cite{Det11}, designed to ease transition from IP, allow also IP routers to cache content. ICN may also exploit caching of routing information to improve its scalability performance \cite{Det12}. Thus, networks of data caches and instruction caches are central in ICN (see e.g. \cite{Psa12,Car11}).

Of course, a model able to predict the performance of cache networks would be important also for more traditional and mature field of applications, such as Content Distribution Networks (CDN) caches, Web caches, etc.

The main contributions of this paper are: i) we evaluate the performance in terms of hit probabilities of a single cache loaded with a renewal input stream; ii) we characterize the miss stream of a single cache in very general terms, modeling the inter-request distribution of such miss stream; iii) we evaluate the performance in terms of hit probabilities of a cache loaded with the miss stream of another cache; iv) we show how such powerful building blocks can be used to evaluate the performance of a network of cache.

Unlike previous work, we do not restrict ourselves to streams modeled with the Independent Reference Model and we do not propose models whose complexity make them useless for designing real world systems.

\section{Related Work}

We limit our description of related work to recent papers devoted to models of multi-cache systems. Exact models for single caches are available (e.g. \cite{Kin71}), but are too complex to be useful for real world applications. Approximate models for single cache (e.g. \cite{Tow90}) and cache hierarchies (e.g. \cite{Che02}) turn out to be more valuable. In particular, \cite{Che02} provides fundamental design principles and useful insights on hierarchical caching. However, exogenous inputs are modeled as Poisson and no explicit model is provided for the output stream, limiting the analysis to a tandem network. Approximate models for arbitrary networks \cite{Ros10} are a significant accomplishment but they are still complex enough to require long processing times and rely on the so-called Independent Reference Model (IRM), to model the streams of requests to {\em all} caches in the network.

The IRM arrival model assumes that ``requests for items occur in an infinite sequence where the item indexes required on the {\em i}-th request, for $i > 0$, are independent random variables on $\{1, 2, \cdots , N\}$ with a common probability distribution'' \cite{Rob12}. Unfortunately this is not true in general both for input exogenous streams and for miss streams of caches directed to other caches, as shown by the authors of \cite{Ros10}.

The work in \cite{Bia13} is the first to our knowledge to model the stream of requests to a cache with a general renewal model instead that with the usual IRM. However, \cite{Bia13} does not consider multi-cache systems.

\section{Model}

We first evaluate the performance of a single cache loaded with a general renewal input stream; then we characterize the miss stream of a single cache; after that we evaluate the performance of a cache loaded with the miss stream of a first-level cache; finally we discuss how to use such powerful building blocks to evaluate the performance of a network of cache.

\subsection{Assumptions and Notation}

In this paper we focus on caches exploiting the Least Recently Used (LRU) replacement policy. We assume, for modeling convenience, that the storage capacity $C$ is expressed as number of items that may be stored therein. This assumption is practical in case of in-network caching, where the cache capacity may be bounded by the size of lookup table used to index the stored items. Otherwise, this assumption implies items of same size; extensions to uneven sizes can be addressed, e.g. as discussed in \cite{Che02, Rob12}.

We assume that items are drawn from an universe size of cardinality $N$. Items are conveniently named using the index $x$, with $x \in \{1,2,...,N\}$. Unlike most past works (e.g. \cite{Tow90,Che02,Rob12}), we characterize the traffic arrival process {\em without} relying on the so-called {\em Independent Reference Model}. Rather, we model the system under more general conditions, by assuming: i) a continuous time scale; ii) inter-arrival times between two consecutive requests for a same item being independent and identically distributed random variables $T_x$, with general cumulative probability distribution function $F_x(t)$ and probability density function $f_x(t)$. When needed, we denote the expected inter-arrival time with $E[T_x] = 1/\lambda_x = \int_{0}^{\infty} \left(1-F_x(t)\right)  \mathrm{d}t$, being thus $\lambda_x$ the {\em average arrival rate} associated to item $x$. We assume stationary arrivals, and consequent long-term item popularity distribution $q_x = \lambda_x / \sum_{i=1}^N \lambda_i$ which, unless otherwise specified, we quantify with a Zipf (non restrictive, as our model does not require to specify any popularity distribution).

We remark that the renewal i.i.d arrival process considered in this paper appears sufficiently descriptive to capture a wide range of \textit{temporal locality} conditions and practical bursty-like traffic patterns, for instance by choosing a random variable $T_x$ with relatively large coefficient of variation \cite{Cro03}.

\subsection{First-level cache with renewal input}

The single-cache model presented in what follows extends, to the renewal input traffic assumption, a clever approximation originally introduced in \cite{Che02} for IRM. Let us focus on an item $x$. The time elapsing between the instant of time the item is inserted (refreshed) in the cache, and the instant of time the item is evicted from the cache (under the assumption than no other requests for the same item arrive to the cache in the meantime), is a random variable with non trivial and a priori unknown distribution.
In \cite{Che02} authors suggest that, for practical (reasonably large) cache sizes and population of items, this random variable can be {\em approximated with a constant}, further {\em independent} of the specific item $x$ considered. This constant is referred to as the ``characteristic time'', $t_c$, of the cache \cite{Che02} and it is a function of the request processes, the cache size, and the request pattern. Despite its simplicity, such an approximation is shown in \cite{Che02} to yield an impressive accuracy, as indeed confirmed by the further analysis provided in \cite{Rob12}.

Under Che's assumption of constant (but unknown) characteristic time, a very simple model can be devised as follows. Indeed, if $t_c$ were known, the probability $H_x$ that a cache hit occurs for item $x\in (1,N)$ would be trivially given by the probability that the inter-arrival time is lower than $t_c$,
\begin{equation}
H_x = P\{ T_x \leq t_c \} = F_x(t_c),
\label{eq:hit}
\end{equation}
resulting in a (weighted) average hit ratio for the whole cache
\begin{equation}
H = \frac{\sum_{x=1}^N \lambda_x H_x}{\sum_{i=1}^N \lambda_i} =
	\sum_{x=1}^N q_x H_x.
\label{eq:hit:tot}
\end{equation}

In order to find $t_c$, we remark that the arrival of a request for an item $x\in (1,N)$ is a {\em renewal} instant. Indeed, irrespective on whether the item was earlier evicted by the cache (and thus the new arrival is a consequence of a cache {\em MISS}, and the item is reinserted) or the item was still in the cache (and thus the new arrival is a {\em HIT}), at the instant of arrival, under the LRU policy, the item history is reset, (the item is logically placed at the {\em top} of the cache), and its future eviction time does not depend on past events, but only on future arrivals. On top of this renewal process, we conveniently define a continuous-time {\em Indicator} process $I_x(t)$, which is equal to 1 when the item $x$ is stored in the cache, and 0 otherwise. From the elementary renewal theorem,
\begin{equation}
E[I_x(t)] \! = \! \frac{E[\mathit{cache \ \! time \ \! per \ \! cycle}]}{E[\mathit{cycle \ duration}]}  \!
=  \! \frac{\int_{0}^{t_c}  \! \left(1 \! - \! F_x(t)\right)  \!  \mathrm{d}t}{1/\lambda_x}
\label{eq:tim}
\end{equation}
where the numerator is the expected value of the random variable defined by $\min(T_x, t_c)$. Indeed, the time spent in the cache in a considered cycle is either the inter-arrival time of the next request for $x$, if this comes before the characteristic time $t_c$, or it is bounded by $t_c$.
The unknown constant $t_c$ can now be computed by imposing the condition that, at each time instant, the cache must contain exactly $C$ distinct items, i.e.,
\begin{equation}
\sum_{x=1}^N I_x(t) = C  \ \ \ \rightarrow \ \ \ \sum_{x=1}^N E[I_x(t)] = C
\label{eq:sum}
\end{equation}

\subsection{Characterizing the cache output stream}

An interesting remark in \cite{Che02} is that a cache can be viewed as a low-pass filter with a cutoff frequency equal to the inverse of the characteristic time of the cache, $t_c$. Here, filtering must be understood in the sense that requests of an item occurring with a frequency lower than $1/t_c$ will result in a cache miss and thus contribute to the {\em miss stream}. Higher frequency requests will find the item in the cache and will not be forwarded to the next cache. Since our model (\ref{eq:hit}) is fully described by the inter-arrival distribution of requests for each item, we can harness and further exploit such a filtering analogy. Indeed, we can look at this filtering process on the time axis: if an arrival of a request occurs later than $t_c$ from the previous one, it will ``pass through'' the cache and contribute to the miss stream; otherwise it will be filtered out.

More specifically, let ${\bar T_x}$ be the r.v. describing the inter-arrival time between two consecutive {\em cache misses} for a same item $x$, and recall that $F_x(t)$ and $f_x(t)$ are the CDF and PDF of the original inter-arrival process $T_x$ offered to the first level cache. By construction, a cache miss is caused by a (last) inter-arrival $T_x > t_c$, possibly preceded by $0$ or more inter-arrival times shorter than $t_c$ (hence filtered out as first level cache hits). Hence, the PDF $f_{\bar x}(t)$ of the r.v. ${\bar T_x}$ can be expressed as (for $t>t_c$, otherwise zero):
\begin{equation}
f_{\bar x}(t) = u_1(t-t_c) f_x(t) * \sum_{k=0}^\infty \left\{ \left(1-u_1(t-t_c) \right) f_x(t) \right\}^{*k}
\label{eq:conv}
\end{equation}
where $u_1(t)$ is the unit-step function, the operator * denotes convolution, $\{g(t)\}^{*k}$ is the n-fold convolution of the (generic) function $g(t)$ with itself, with the usual convention that $\left\{g(t)\right\}^{*0}$ yields the Dirac $\delta(t)$.

It is also useful to derive compact expressions for mean and variance of ${\bar T_x}$. For the mean value, there is no need to perform any computation (although for completeness a direct derivation is provided in the Appendix), as it suffices to derive $E[{\bar T_x}]$ as the inverse of the miss stream frequency, i.e.,
\begin{equation}
E[{\bar T_x}] = \frac{1}{\lambda_x (1-H_x)} = \frac{E[T_x]}{1-H_x}.
\label{e:mean}
\end{equation}
The variance instead requires some more algebra (see Appendix) and is expressed in terms of the statistics of the original inter-arrival time $T_x$ by
\begin{equation}
Var[{\bar T_x}] = \frac{Var[T_{x}]}{1-H_x} - \frac{H_x ( E[T_{x}]^2 - 2 E[T_x] E[T_x | T_x \leq t_c])}{(1-H_x)^2}
\label{e:var}
\end{equation}
Finally, dividing (\ref{e:var}) by the square of (\ref{e:mean}) yields the square of the Coefficient of Variation
\begin{equation}
C_{\bar x}^2 = \frac{Var[{\bar T_x}]}{E[{\bar T_x}]^2} = C_x^2 (1-H_x) + H_x \left(
	\frac{2 E[T_x | T_x \leq t_c]}{E[T_x]} -1 \right)
\label{e:cv2}
\end{equation}
Note that this value depends on the cache filtering effect. For instance, if we assume exponentially distributed inter-arrival of requests as input, (\ref{e:cv2}) simplifies to $1-2 \lambda_x t_c e^{-\lambda_x t_c}$, showing that the cache has a varying smoothing effect on the CV, depending on the product $\lambda_x t_c$, with smoothing maximum at $\lambda_x t_c = 1$, with a resulting $CV=\sqrt{1-2/e}$.

\subsection{Second-level cache}
At this point it is easy to evaluate the performance of a cache loaded solely with the output stream of another cache. To derive the probability that a cache hit occurs for item $x\in (1,N)$ at the second-level cache it suffices to apply (\ref{eq:hit}) using now for the probability density function of the inter-arrival times between two consecutive requests for a same item (\ref{eq:conv}).

\subsection{Cache networks}
The arrival stream to a generic cache in a network derives from a superposition of exogenous inputs and of miss streams of other caches. If we assume independence between such components it is relatively easy to evaluate the overall probability density function of the inter-arrival times between two consecutive requests for a same item by means of convolutions and then apply (\ref{eq:hit}).
However, the power of the approximation by Che et. al. \cite{Che02} is such that more compact and expedient expressions can be derived, which we leave for further work.

\section{Numerical Results}

\begin{figure}[t]
\includegraphics[width=8.5cm, height=4.3cm]{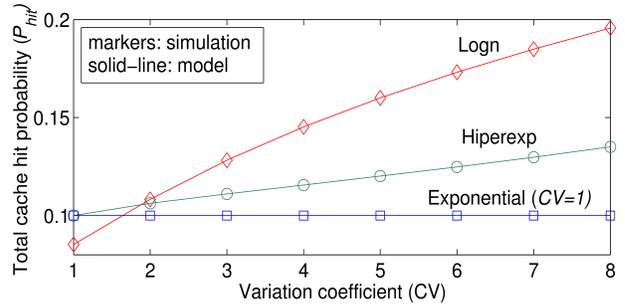}
\caption{Total cache hit probability vs. variation coefficient}
\centering
\label{fig:hit_cv}
\end{figure}

We start with the analysis of a first-level cache. Figure \ref{fig:hit_cv} shows the total cache hit probability for three different distributions of the request inter-arrival time (Exponential, Hyper-Exponential, and LogNormal). For the case of Hyper-Exponential and LogNormal, different coefficients of variations $CV$, ranging from 1 to 8, are plotted. Model (\ref{eq:hit:tot}) and simulations are compared, showing a perfect fit. Performance significantly depend on the chosen inter-request distribution and improve for a greater CV.

\begin{figure}[t]
\includegraphics[width=8.5cm, height=4.5cm]{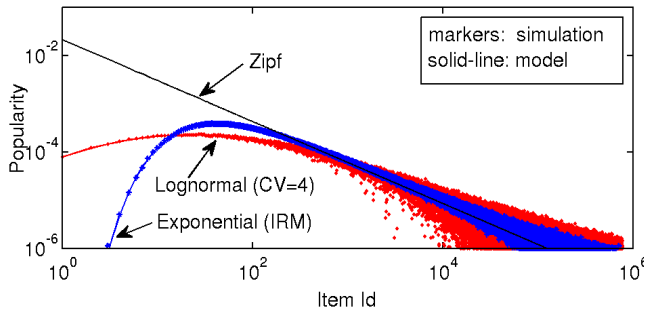}
\caption{Miss stream popularity distribution}
\centering
\label{fig:pop}
\end{figure}

Turning now to second-level cache issues, Figure \ref{fig:pop} shows how the Zipf popularity law of requests arriving to the first cache $q_x$ is modified by the first-level cache: the popularity distribution of the miss stream $q'_x = q_x (1 - H_x)$, shown for both exponential and lognormal distribution of the inter-arrival process of requests at the first cache, is a filtered replica of the ingress one; the miss stream popularity $q'_x$ computed using the hit probability $H_x$ with our model (\ref{eq:hit}) closely follows simulations.

\begin{figure}[t]
\includegraphics[width=8.5cm]{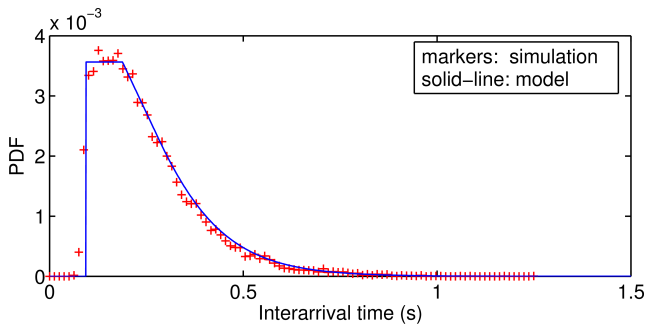}
\caption{Probability density function (PDF) of an item on the miss stream}
\centering
\label{fig:item_pdf}
\end{figure}

Figure \ref{fig:item_pdf} shows the probability density function of the inter-arrival time of the most popular item (i.e., the first item of the input stream, with a first-level cache of size $100$) in the miss stream, comparing simulations and model (\ref{eq:conv}); the inter-arrival process of item requests to the first-level cache is exponential. The filtering effect is evident: the pdf is zero for inter-arrival times less than $t_c$ (equal to $93.2ms$). Once again the fitting is remarkable.

\begin{figure}[ht!]
     \begin{center}
        \subfigure[Exponential input stream]{
            \label{fig:exp_pdf}
            \includegraphics[width=8.5cm]{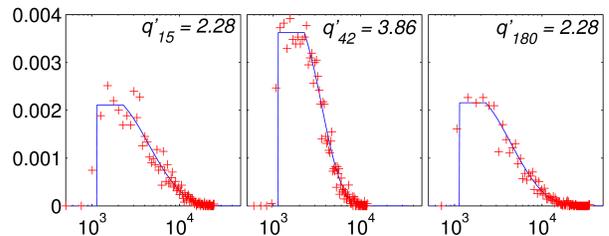}
        }\\
        \subfigure[Lognormal input stream]{
           \label{fig:logn_pdf}
           \includegraphics[width=8.5cm]{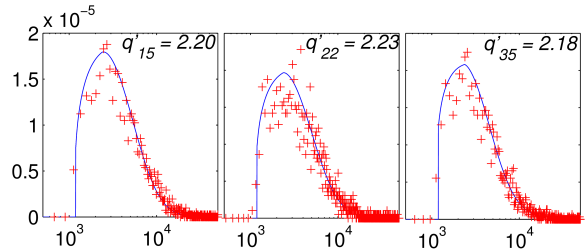}
        }\\
    \end{center}
    \caption{PDF of the output (miss) stream for different items; $q_x' * 10^{-4}$ is the (normalized) popularity at the output of the cache (x-axis in ms)}
   \label{fig:miss_pdf}
\end{figure}

Figure \ref{fig:miss_pdf} shows the same performance measure for other items and for both exponential and lognormal distribution of the inter-arrival process at the first cache.

\begin{figure}[t]
\includegraphics[width=8.5cm]{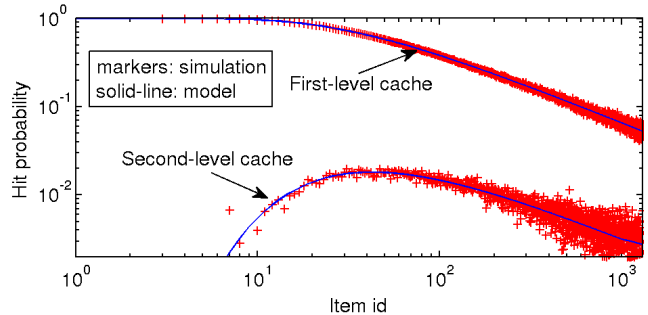}
\caption{Per item hit-rate on the second cache for exponential distribution on the first cache}
\centering
\label{fig:exp_hit_both}
\end{figure}

Finally, Figure \ref{fig:exp_hit_both} shows how our model performs in evaluating the item hit-rate on the second cache, with an exponential distribution of the inter-arrival process at the first cache. Also plotted is the hit probability of the first-level cache. The difference between the two is apparent and confirms the need of suitable models for multi-cache systems.

We conclude with three remarks: i) as noted in \cite{Rob12}, the approximation \cite{Che02} works very well even beyond the applicability scenarios stated by its own authors; in addition to the results presented above, we followed a suggestion of \cite{Rob12} and used the approximation \cite{Che02} instead of the model \cite{Tow90} for the single cache approximation used in \cite{Ros10} to evaluate cache networks; the result was a 30th fold decrease of the computing time in some exemplary cases, with a remarkable accuracy;  ii) the summation (\ref{eq:conv}) converges very rapidly: few iterations are enough to reach accuracy in the order of $10^{-6}$ in some exemplary cases; iii) our model for a cascade of caches is general, as it allows using any renewal distribution, tractable as it requires simple algebra with low computing time and accurate, as our results make evident. A more through analysis of these results, together with ensuing design principles and applications, and explicit extension of the model to cache networks is left for further work.

\section{Acknowledgements}
This work is supported in part by the European Commission in the context of the FP7/NICT EU-JAPAN GreenICN project.

\section*{Appendix}

In the following derivations, with slight abuse of notation, it is convenient to define the conditional random variables $T_{x,h} = (T_x | T_x > t_c)$ and $T_{x,l} = (T_x | T_x \leq t_c)$, i.e., random variables defined by the conditional distribution of $T_x$ given $T_x > t_c$ or $T_x \leq t_c$, respectively. $T_{x,h}$ accounts for the inter-arrival time {\em longer} than $t_c$, which causes a cache miss, whereas $T_{x,l}$ accounts for the $0$ or more inter-arrival times {\em shorter} than $t_c$, which are filtered out as first level cache hits. Clearly, owing to truncation at $t_c$, the PDF of $T_{x,h}$ is simply $f_x(t)/(1-H_x)$ for $t>t_c$, whereas the PDF of $T_{x,l}$ is $f_x(t)/H_x$ for $t\leq t_c$.

We note that the r.v. ${\bar T_x}$ describing the inter-arrival time between two consecutive {\em cache misses} for a same item $x$, comprises two parts: i) exactly one inter-arrival time longer than $t_c$, i.e. one r.v. $T_{x,h}$, plus ii) a random number $K \in (0, \infty)$ of r.v.s $T_{x,l}$ accounting for the inter-arrival times shorter than $t_c$. $K$ is a geometrically distributed r.v. with probability distribution $P(K=i) = (1-H_x) H_x^i$ and thus mean value $E[K] = H_x/(1-H_x)$ and variance $Var[K] = H_x/(1-H_x)^2$.

{\bf \em Direct derivation of $E[{\bar T_x}]$}. From elementary conditional expectation arguments,
\[ E[{\bar T_x}] = E[T_{x,h}] + E[K] E[T_{x,l}]  = \]
\[ = \int_{t_c^+}^{\infty}  \! \frac{t f_x(t)}{1\!-\!H_x}    \mathrm{d}t +
	\frac{H_x}{1\!-\!H_x}
	\int_{0}^{t_c}  \! \! \frac{t f_x(t)}{H_x}   \mathrm{d}t = \frac{E[T_x]}{1-H_x}. \]

{\bf \em Derivation of $Var[{\bar T_x}]$}. We make use of well known results on the variance of the sum of a random number $K$ of r.v.s, and we exploit independence among the involved r.v.s. Therefore,
\[ Var[{\bar T_x}] = Var[T_{x,h}] + E[K] Var[T_{x,l}] + Var[K] E[T_{x,l}]^2  =\]
\begin{equation}
= E[T_{x,h}^2]-E[T_{x,h}]^2 + \frac{H_x (E[T_{x,l}^2]\!-\!E[T_{x,l}]^2)}{1-H_x} +
 \frac{H_x E[T_{x,l}]^2}{(1-H_x)^2}
\label{eq:var1}
\end{equation}
We then exploit the relations involving the original inter-arrival time $T_x$ and the related conditional random variables:
\[ E[T_{x}]= (1-H_x) E[T_{x,h}] + H_x E[T_{x,l}] \]
\[ E[T_{x}^2]= (1-H_x) E[T_{x,h}^2] + H_x E[T_{x,l}^2] \]
Equation (\ref{eq:var1}) hence simplifies to
\[ Var[{\bar T_x}] = \frac{E[T_{x}^2]}{1-H_x} + \frac{H_x^2 E[T_{x,l}]^2-(1-H_x)^2 E[T_{x,h}]^2}{(1-H_x)^2} =  \]
\[ = \frac{E[T_{x}^2]}{1-H_x} + \frac{2 H_x E[T_x] E[T_{x,l}]- E[T_{x}]^2}{(1-H_x)^2} = \]
\begin{equation}
= \frac{Var[T_{x}]}{1-H_x} - \frac{H_x }{(1-H_x)^2}  \left( E[T_{x}]^2 - 2 E[T_x] E[T_{x,l}] \right)
\end{equation}

\balance

\bibliographystyle{abbrv}
\bibliography{ref}

\end{document}